\documentclass[twocolumn]{article}
\usepackage{graphicx,amsmath,amssymb,color,cite}
\usepackage{hyperref}
\hypersetup{
	colorlinks,
	linkcolor={blue},
	citecolor={blue},
	urlcolor={blue!80!black}
}

\begin{document}

\title{Balancing a Stick with Eyes Shut: Inverted Pendulum on a Cart without Angle Measurement}
\author{Bidhayak Goswami\thanks{bidhayak1728@gmail.com,bidhayak@iitk.ac.in} \and Anindya Chatterjee\thanks{anindya100@gmail.com, anindya@iitk.ac.in}}

\date{Mechanical Engineering, IIT Kanpur\\ \smallskip \today
\\\smallskip{\em A shorter version of this paper is due to appear in ASME JDSMC.} }

\maketitle

\begin{abstract}
We consider linear time-invariant dynamic systems in the single-input, single-output (SISO) framework. In particular, we consider stabilization of an inverted pendulum on a cart using a force on the cart. This system is easy to stabilize with pendulum angle feedback. However, with cart position feedback it {\em cannot} be stabilized with stable and proper compensators. Here we demonstrate that with an additional compensator in a parallel feedforward loop, stabilization is possible with such compensators. Sensitivity to noise seems to be about 3 times worse than for the situation with angle feedback. For completeness, discussion is presented of compensator parameter choices, robustness, fragility and comparison with another control approach.
\end{abstract}


\section{Introduction}
Stabilization of an inverted pendulum on a cart is a familiar problem in control theory, and also one that is interesting to a broad audience. The control input is a horizontal force on the cart; and it is desired to use feedback to stabilize the cart at a given location and the pendulum in the inverted (or standing vertical) position. Formally, the linearized system is controllable. In the classical control theory framework, with a single input and a single output (SISO), it is important whether the output is the pendulum angle or the cart position. Due to the obvious resemblance to balancing a stick on one's palm, we refer to the latter case (i.e., cart position known and pendulum angle unknown) as {\em balancing a stick with eyes shut}. This problem, although simple to state, is interesting to a broad audience because a few attempts to balance a stick on one's palm with eyes shut will convince the reader that the task is difficult if not impossible.
Technically, the problem is also interesting within the usual classical single-loop feedback control framework because, in the case of solely position feedback, it turns out that the system is {\em not stabilizable with a stable and proper controller} \cite{youla1974single}.

As a control problem, the inverted pendulum is a familiar favorite. It has been studied by several researchers in the past \cite{blitzer1965inverted,phelps1965analytical,kalmus1970inverted,mori1976control}. It has been used as a teaching example for many decades \cite{cannon2003dynamics,richard2008modern}.
Linear control theory has been used for the angular position feedback case, and that case does not represent significant challenges any more. Control in the nonlinear regime, including swing-up dynamics from a hanging-down position, has been studied \cite{furuta1992swing,bradshaw1996swing,aastrom2000swinging,bugeja2003non,siuka2009applications} from various viewpoints including energy-based control as well as control input shaping.
Some researchers have studied the effectiveness of simple feedback laws with delays \cite{milton2009time}, again with angular position feedback. Lee et al.\ \cite{lee2015output} have considered system uncertainties, feedback with multi-timescale structure and an extended high-gain observer.
An optimal control approach has been used as well, even for harder variants, e.g., a double inverted pendulum on a moving cart \cite{bogdanov2004optimal}. 
With advances in control theory, more modern techniques like robust control, fuzzy logic control, etc.\ have been considered as well \cite{dastranj2012robust,nour2007fuzzy,zabihifar2020robust}.

In spite of the abovementioned works with modern approaches, in this paper we remain within the classical SISO linear regime for two reasons. Firstly, a large number of industrial control systems are still linear in their thinking, and often close to just PID control or variations thereof. Secondly, in the absence of angle feedback, for the inverted pendulum on a cart, we obtain a nonminimum phase system with an odd number of real poles on the right of an RHP real zero which, it is known, cannot be stabilized by a stable and proper compensator in the usual single feedback loop configuration. 

Of course, not all control is based solely on feedback. Feedforward compensators \cite{bar1987parallel,palis2014discrepancy} have been used earlier for stabilizing nonminimum phase systems. Kim et al.\ \cite{kim2016design} used feedforward compensation for the synchronization of a multi-agent system to achieve faster convergence. Golovin and Palis \cite{golovin2017design,golovin2021pfc} used feedforward compensation for stabilizing an electromechanical system with friction-induced instabilities. Here, we seek suitable feedforward and feedback compensators, {\em both stable and proper}, to stabilize the inverted pendulum using only the cart position as output.

In what follows, we describe the system in section \ref{system_and_eq}, present the control approach and show some relevant results in section \ref{feedforward_ctrl}, and discuss the effect of noise in section \ref{effect_of_noise}. A detailed derivation of the equations of motion is given in appendix \ref{app_eqn_motion}, a description of the controller design methodology is given in appendix \ref{method}, the modern control approach in terms of controllability and observability is discussed in appendix \ref{app_modern_control}, robustness and fragility of the stable compensators have been examined in appendix \ref{robustness}, and finally, in appendix \ref{app_noise}, the effect of noise on the system has been discussed.

\section{The inverted pendulum}\label{system_and_eq}
\begin{figure}[h!]
	\centering
	\includegraphics[scale=0.4]{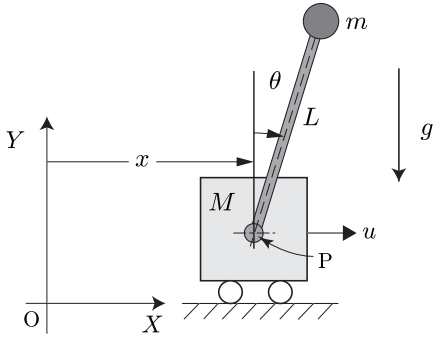}
	\caption{An inverted pendulum on a cart.}
	\label{inv_pend}
\end{figure}

Consider an inverted pendulum on a cart (Fig.\ \ref{inv_pend}). A control force $u$ acts on the cart as shown. The pendulum consists of a massless rigid rod of length $L$ with a point mass $m$ at the tip. The cart's mass is $M$. There is gravity $g$. The coordinates used are $\theta$ for the pendulum angle and $x$ for the cart displacement. 

The equations of motion (see appendix \ref{app_eqn_motion} for details) for small $\theta$ and $\dot \theta$ are
\begin{subequations}
	\begin{equation}
		\label{eqm1}
		M \ddot x + m\left ( \ddot x + L \ddot \theta \right ) = u,
	\end{equation}
	\begin{equation}
		\label{eqm2}
		\ddot x + L \ddot \theta = g \theta. 
	\end{equation}
\end{subequations}
By choice of units of mass, length and time, we can set $m$, $L$ and $g$ to unity. This leaves
\begin{subequations}
	\begin{equation}
		\label{meqm1}  
		M\ddot x +  \left ( \ddot x + \ddot \theta \right ) = u,
	\end{equation}
	\begin{equation}
		\label{meqm2}
		\ddot x + \ddot \theta = \theta,
	\end{equation}
\end{subequations}
where {$M$} should henceforth be thought of as dimensionless.

\begin{figure}[h!]
	\centering\includegraphics[width=\linewidth]{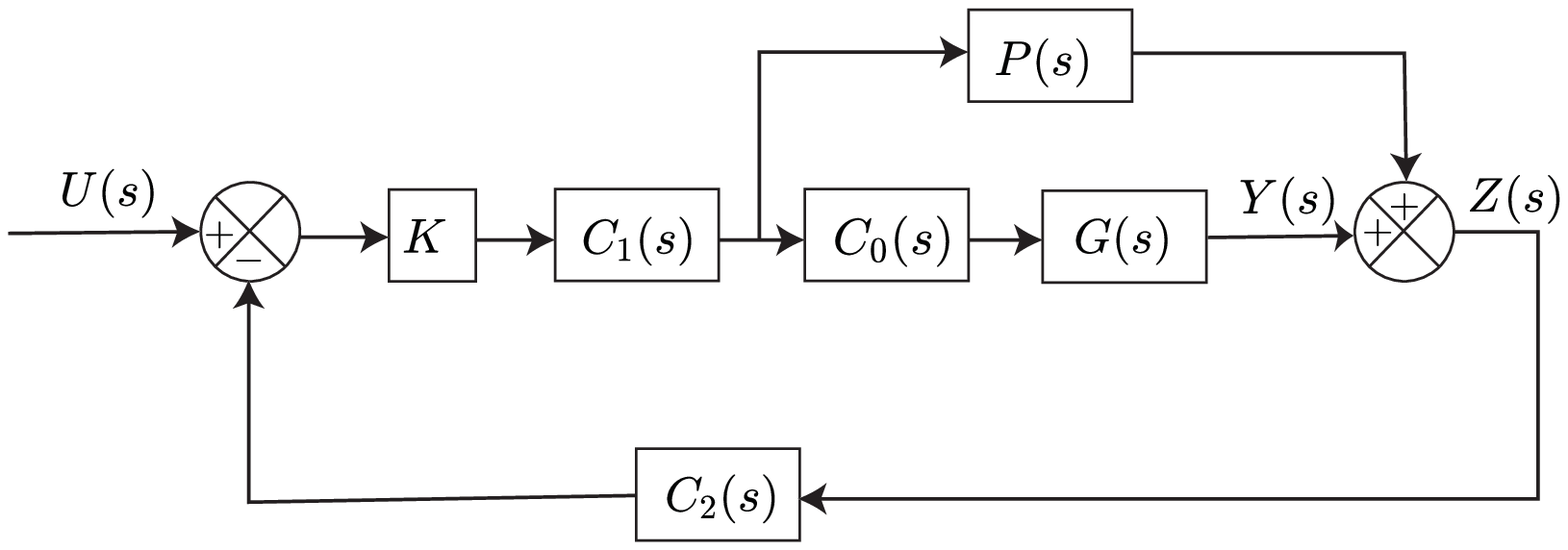}
	\caption{A feedfoward-feedback control system: basic layout. We take \(C_0=1\) for simplicity (see section \ref{feedforward_ctrl} for justification).}
	\label{tf}
\end{figure}

In the SISO framework within elementary classical control, we have a single output: this will often be either the pendulum angle or the cart position.
See the block diagram in Fig.\ \ref{tf} for the basic control structure we will consider in this paper. Here, $U$ is the input, $K$ is a gain, $G$ is the plant, {$ C_0 $}, $C_1$ and $C_2$ are compensators, $P$ is an optional feedforward compensator placed in parallel, $Y$ is the actual plant output, and $Z$ is the quantity that is fed back. Obviously, in the absence of $P$, or with $P=0$, we will have $Z=Y$ and obtain the usual single-loop feedback control design. The parallel
feedforward compensator $P$ is the novelty we will consider here.

A feedforward compensator before the input signal \(U\) is routinely used in physical systems where the input from the user may be, e.g., a desired angle and the input to the measurement and control system is, typically, a voltage. However, subsequent analysis of the control system is independent of that compensator. So, we have not included it in this paper.

Returning to the inverted pendulum on a cart, when we balance a stick on our palm, we always look at the pendulum angle $\theta$. The corresponding transfer function of the plant is
\begin{equation}
\label{G1}
F=\frac{1}{1+M-M\,s^2},
\end{equation}
which has a right half plane or RHP pole but is easily stabilizable with a stable controller without any $P$ in a parallel feedforward path.
For example, with $M=0.3$, $C_0=C_1=1$, and $K>4.33$, 
\begin{equation}\label{F_val}
F=\frac{1}{1.3-0.3 s^2}
\end{equation}
 can be stabilized by the stable compensator
\begin{equation}
\label{C2}
C_{2} = - \frac{s+3}{s+10}.
\end{equation}

However, if our output variable is $x$, which loosely corresponds to trying to balance the stick on our palm with our eyes shut, then the
plant becomes
\begin{equation}
	\label{e2}
	G=\frac{s^2-1}{s^2 \left (0.3 s^2 - 1.3 \right)},
\end{equation}
which has a real RHP zero at $s=1$ and a single real RHP pole to its right, at $s=\sqrt{1.3/0.3}$. This is more interesting.

\section{Stabilization}\label{feedforward_ctrl}

In classical control theory with a single control loop \cite{ogata2010modern},  $P=0$. Then, although \(C_0\), $C_1$ and $C_2$ can in principle all be present, stability is affected only by the product $C_0C_1C_2$, and so we can for stabilization purposes take any two of them to be unity. The gain $K$, too, can be included within $C_1$ if we wish.
In this paper, we take \(C_0=1\) for simplicity. From a design viewpoint, non unity \(C_0\) can be thought as a system modification and we leave it for more challenging problems.

In simple control systems, the compensators $C_1$ and $C_2$ may be physically realized using simple circuits made with resistors, capacitors, and op-amps. In such cases we want the compensator transfer functions themselves to be stable, i.e., \(C_0\), $C_1$ and $C_2$ should not have RHP poles. Here we assume that $P$ has no RHP poles either. Thus we are interested in stabilizing $G$ with stable controllers\footnote{%
	The motivation is that the analog control card should not saturate and lose linearity before the actual system dynamics is established, e.g.,
	during a warm-up phase.}.

In the absence of $P$, a fundamental fact has been known for almost 50 years \cite{youla1974single}.
If $G$ has one or more real RHP zeros; and if $G$ also has an odd number of real RHP poles that lie to the right of any positive real zero; then in the absence of $P$ in Fig.\ \ref{tf}, $G$ is not stabilizable \cite{youla1974single} with stable and proper compensators \(C_0\), $C_1$ and $C_2$.

The mathematical aim of this paper can now be clearly stated. We will take the troublesome $G$ of Eq.\ (\ref {e2}),  set $K=C_1=1$ (as well as \(C_0=1\) as stated above), and find a stable and proper $C_2 = C$ along with a stable and proper $P$ such that
the plant output $Y$ is stabilized.

Let
$$ C = \frac{n_C}{d_C}, \quad G=\frac{n_G}{d_G},\quad \mbox{ and } P = \frac{n_P}{d_P},$$
where the $n$'s stand for numerator polynomials in $s$ and the $d$'s stand for denominator polynomials in $s$ of equal or greater order.
Routine manipulations show that
\begin{equation}\label{Y_tf}
	Y = \frac{GU}{1+C(P+G)} = H U,\end{equation}
i.e., the controlled transfer function is
\begin{equation}\label{CLTF}
H =  \frac{n_G d_C d_P}{d_C d_G d_P+n_C n_P d_G
+n_C n_G d_P}.
\end{equation}

Thus, our controller design for stabilization reduces to choosing polynomials $n_C$, $d_C$, $n_P$ and $d_P$ such that  the $d$-polynomials are stable (i.e., they have only LHP roots), and the denominator polynomial
$$ d_C d_G d_P+n_C n_P d_G
+n_C n_G d_P$$
is stable as well (has only LHP roots).

We are not aware of systematic and guaranteed ways of obtaining such polynomials. We have used trial and error based on a simple optimization routine.
Some details of the optimization are given in appendix \ref{method}. Our main aim here is to demonstrate and assess specific numerical solutions.

\begin{figure}[h!]
	\centering
	\includegraphics[width=\linewidth]{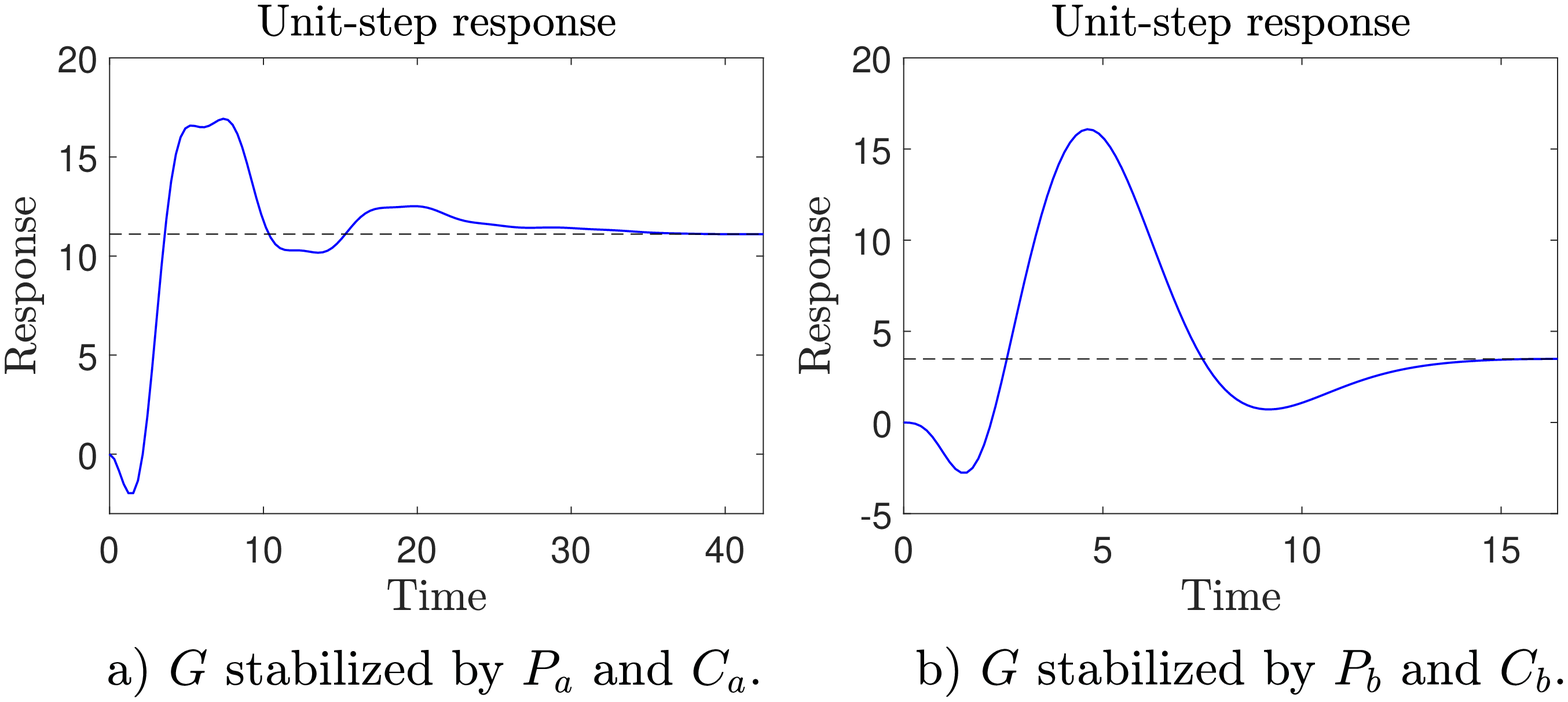}
	\caption{Unit-step responses of  $ G $ of Eq.\ (\ref{e2}), with $ P $ and $ C $ as given by Eq.\ (\ref{P_C_2}) and Eq.\ (\ref{P_C_2a}).}
	\label{step_2}
\end{figure}

Two stabilizing solutions, out of many that we found, are shown below in Eqs.\ (\ref{P_C_2}) and (\ref{P_C_2a}), labeled ``a'' and ``b'' respectively.
\begin{align}\label{P_C_2}
	P_a &=  \frac{0.05 s^3  +  7 s^2   - 0.1 s   -1.9}{11 s^3  + 21.7 s^2   + 5.4 s + 1}
	, \nonumber\\
	  C_a&= \frac{-10.1 s^3 +2 s^2+   0.9s+    0.09 }{  0.002 s^3  +4.2s^2  +10.2s + 1},
\end{align}
and
\begin{align}\label{P_C_2a}
	P_b &= \frac{ 0.2 s^3 + 1.4 s^2 - 2 s - 0.8 } { 4.1 s^3 + 10.8  s^2 + 5.3 s + 1 },\nonumber \\
	C_b& = \frac{ -6.9 s^3 + 1.6 s^2 + 1.1 s + 0.3}{ 0.08 s^3 + 0.4 s^2 + 9.3 s + 1 }.
\end{align}
The corresponding unit-step responses of the controlled systems are shown in Fig.\ \ref{step_2}.

\begin{figure}[h!]
	\centering
	\includegraphics[width=\linewidth]{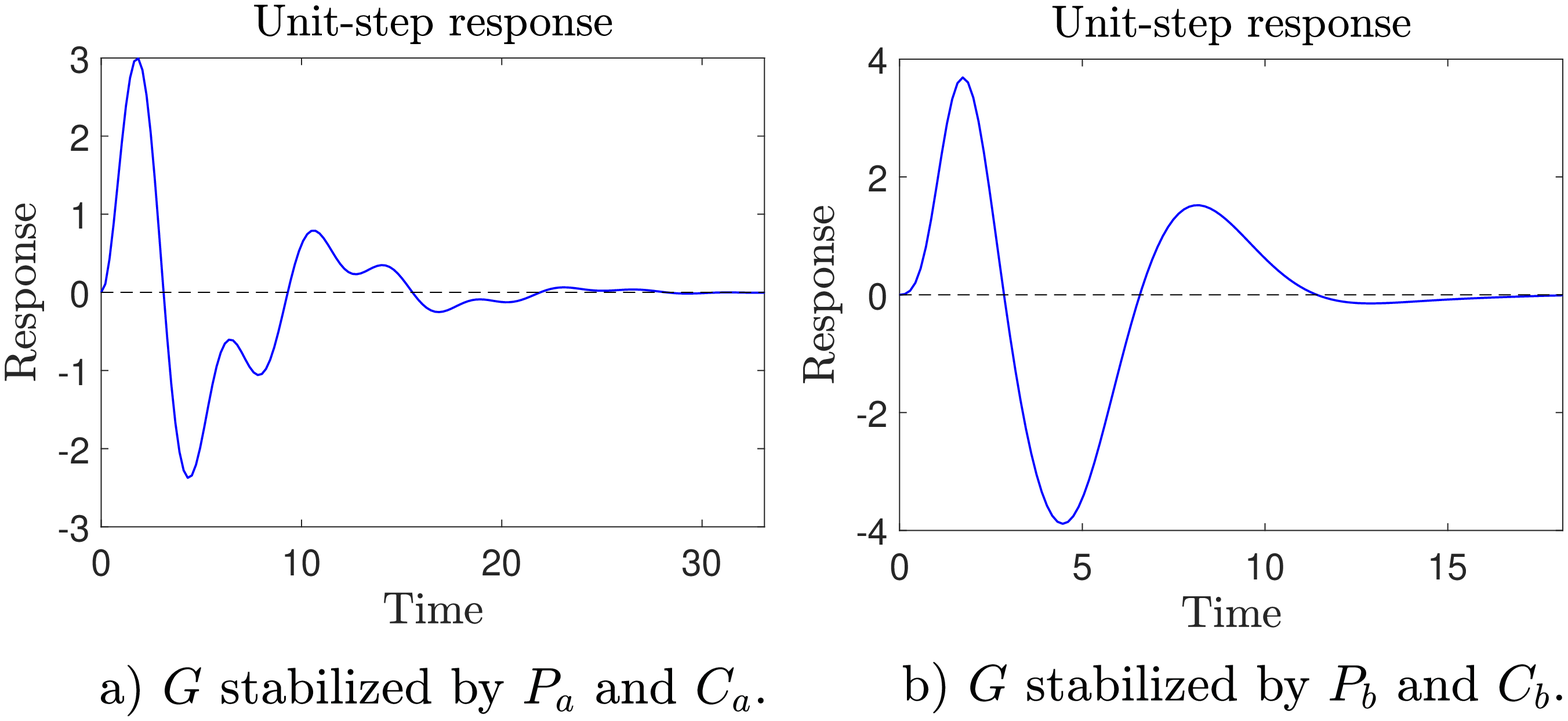}
	\caption{Angular response of the pendulum.}
	\label{angres}
\end{figure}

At this point we can check to see, for the same controlled system (with only position feedback), the angle response of the inverted pendulum. Recalling $F$ from Eq.\ (\ref{F_val}), we write
$$ n_F=1,\quad \mbox{and}\quad d_F=1.3-0.3 \,s^2.$$
Further, recalling Eqs.\ (\ref{e2}), (\ref{Y_tf}) and (\ref{CLTF}), we find that the angular response of the inverted pendulum must be
$$\frac{HF}{G} \,U ,$$
with 
\begin{equation}\label{Angle_tf}
\frac{HF}{G}=-\frac{n_F d_C d_P}{d_C d_G d_P+n_C n_P d_G
	+n_C n_G d_P}\,s^2,
\end{equation}
where we have used
$$ d_G=-s^2\,d_F.$$
The angular response of the pendulum is given by the step response of Eq.\ (\ref{Angle_tf}): see figure \ref{angres}.

It is also interesting to ask how other control strategies might work for this same system. A discussion of the textbook approach of modern control theory, with
state estimation and full state feedback, is given in appendix {\ref{app_modern_control}}.

Finally, we must address two more issues: (i) the robustness of the controller, i.e., its ability to stabilize plants with slightly different plant-parameter values, and
(ii) the fragility of the controller, i.e., its tendency to lose effectiveness under small perturbations of the controller-parameter values. Both robustness and fragility are good, as shown in appendix \ref{robustness}.

\section{Effect of noise}\label{effect_of_noise}
The system has now been mathematically stabilized. We must also study the effect of noise on the stabilized system. Since the basic control problem is difficult
at least in some ways (as in balancing a stick on one's palm with eyes shut), we may expect increased sensitivity to noise.
\begin{figure}[h!]
	\centering
	\includegraphics[scale=0.6]{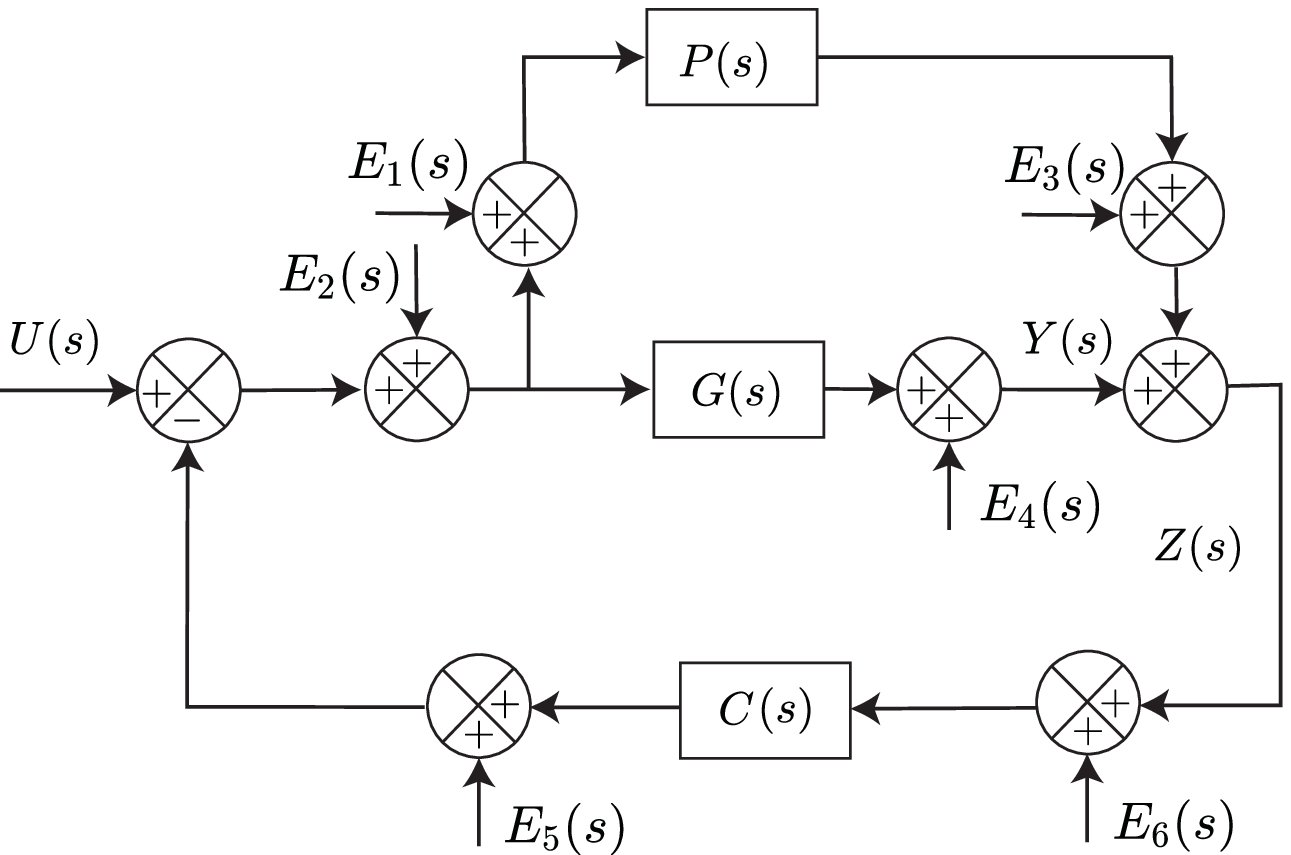}
	\caption{The control system with added noise inputs $ e_k(t),\,k\in\{1,2,\dots,6\} $.}
	\label{sys_with_noise}
\end{figure}
The system now has more than one input (the control force $ u $ along with noise inputs $ e_k,\,k\in\{1,2,\dots,6\} $), but still only one output ($ y $), as shown in Fig.\ \ref{sys_with_noise}. In the Laplace domain, elementary calculations give us individual transfer functions for each input, with the net output given as
\begin{equation}
	Y(s) = H(s) U(s) + \sum_{k=1}^{6}H_{{ e}_k}(s) E_k(s). 
\end{equation}
In the above the seven transfer functions $ H $ and $ H_{ e_k},\,k\in\{1,2,\dots,6\}, $ share a common denominator. If $ H $ is stable, then so is each $H _{ e_k} $. The Bode magnitude plots for $ H_{ e_k} $ with $ P=P_b $ and $ C=C_b $ are shown in Fig.\ \ref{fig_noise}. For $P_a$ and $C_a$, the maximum magnitude is higher. It is seen in Fig.\ \ref{fig_noise} that for nondimensional frequencies on the order of unity, the magnitudes of the transfer functions take their highest values, which are around 30 dB. This corresponds to amplification by roughly a factor of 30, which is quite large.
\begin{figure}[h!]
	\centering
	\includegraphics[width=\linewidth]{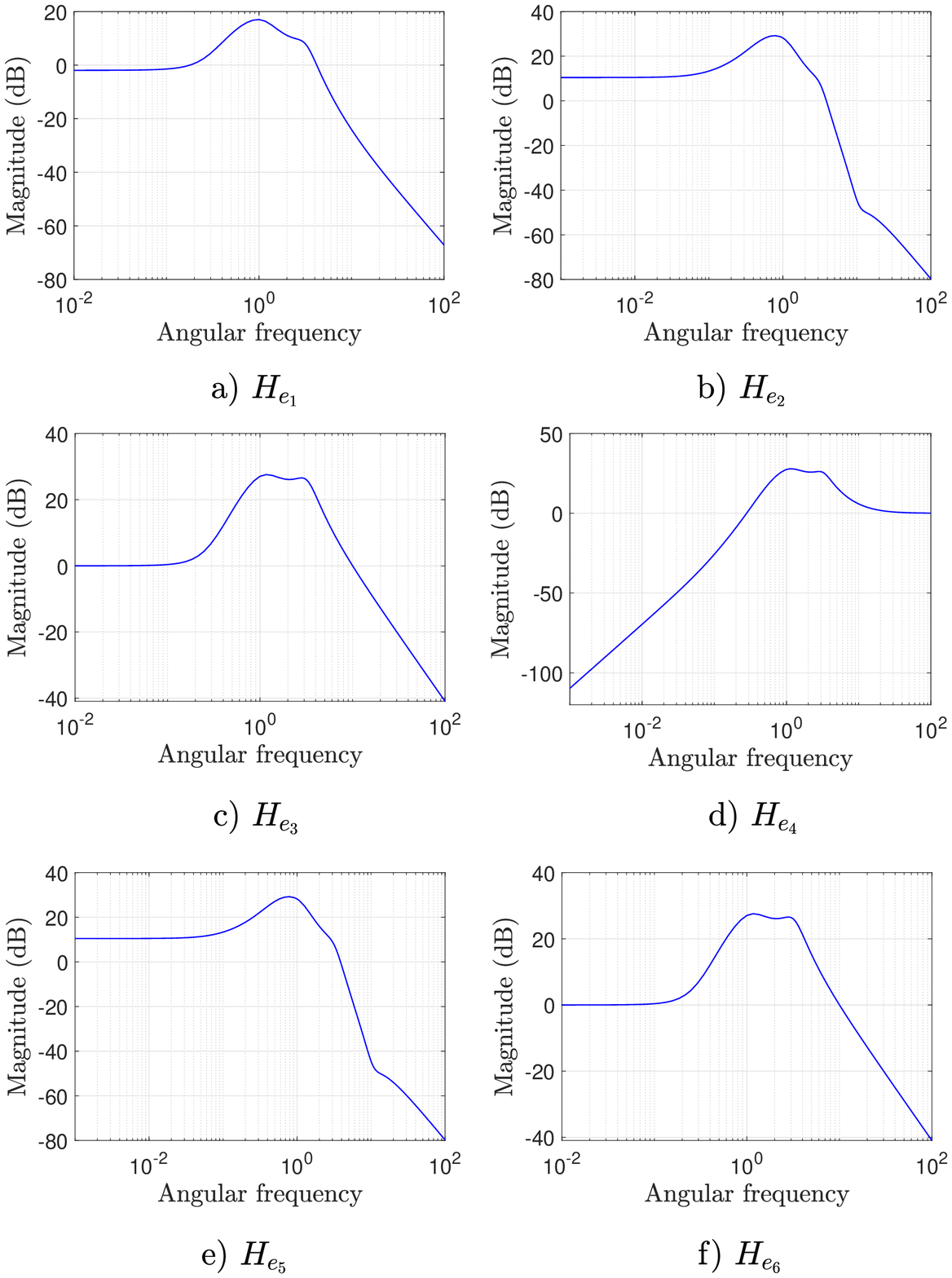}
	\caption{Bode magnitude plots of $ H_{e_k},\,k\in\{1,2,\dots,6\}, $ for the system shown in Fig.\ \ref{sys_with_noise} with $P=P_b$ and $C=C_b$ given by Eq.\ (\ref{P_C_2a}).}
	\label{fig_noise}
\end{figure}

We mention that in separate calculations with $G$ as in Eq.\ (\ref{G1}), $K>4.33$, $P=0$ (i.e., no parallel feedforward compensator), and $C=C_2$ of
Eq.\ (\ref{C2}), maximum amplification factors about 10 dB lower were easily obtained (details omitted). This is not intuitively
surprising because balancing a stick with one's eyes open is easier than with one's eyes shut; and correspondingly, stabilizing the inverted pendulum on a cart in the classical SISO setting with pendulum angle feedback is easier than with cart position feedback. The sensitivity to noise in the eyes-shut case, for our control design, seems to be greater by about a factor of 3.

Some numerical examples of the response to noise inputs, where the ``noise'' is the sum of a large number of small sinusoidal inputs with randomly chosen amplitudes and frequencies, are given in appendix \ref{app_noise}. The results there are consistent with the above estimate of 30 dB.

\section{Concluding remarks}
In some applications such as low cost consumer products or toys, there may be simple analog control cards which, if operated within the linear domain, produce desirable behavior in the controlled device. In such situations, stabilization with a stable controller has practical utility. Additionally, there may be sophisticated scientific or technical instruments wherein simple controllers are implemented with some parameters adjustable by the user. In such cases, too, while the system is warming up, or is outside the operational position range, an unstable compensator may lead to overly large actuator commands that cause saturation, deviation from linearity, or possibly specimen damage. In such cases, also, stabilization with a stable controller may make the system more foolproof.

With the above motivation, we appreciate the classic paper \cite{youla1974single} which lays down the mathematical conditions under which, in the single loop control structure, stabilization is not possible with a stable and proper compensator. One of the most familiar control problems, namely balancing an inverted pendulum on a cart, presents this situation when position feedback is used. For this system, using a parallel stable feedforward compensator, we have demonstrated using numerical examples that it is possible to achieve stabilization with stable and proper compensators.

Finding such stable and stabilizing compensators is not a familiar and routine control design problem within classical control theory. Others have studied this control approach before as well \cite{kim2016design,golovin2017design}, but not widely and not for such a popular problem as balancing a stick.
We hope that the community of industrial and academic control systems practitioners and researchers will find this class of problems sufficiently interesting
as to develop this kind of control design further, and possibly even seek rational design criteria for when such controllers exist and how they may be found easily.

Moreover, once such a block diagram framework is adopted, it can also be used for design of controllers for nonlinear systems. Such work \cite{robenack2019control} has begun to appear.

\appendix

\section{Equations of motion}\label{app_eqn_motion}
\begin{figure}[h]
	\centering
	\includegraphics[scale=0.48]{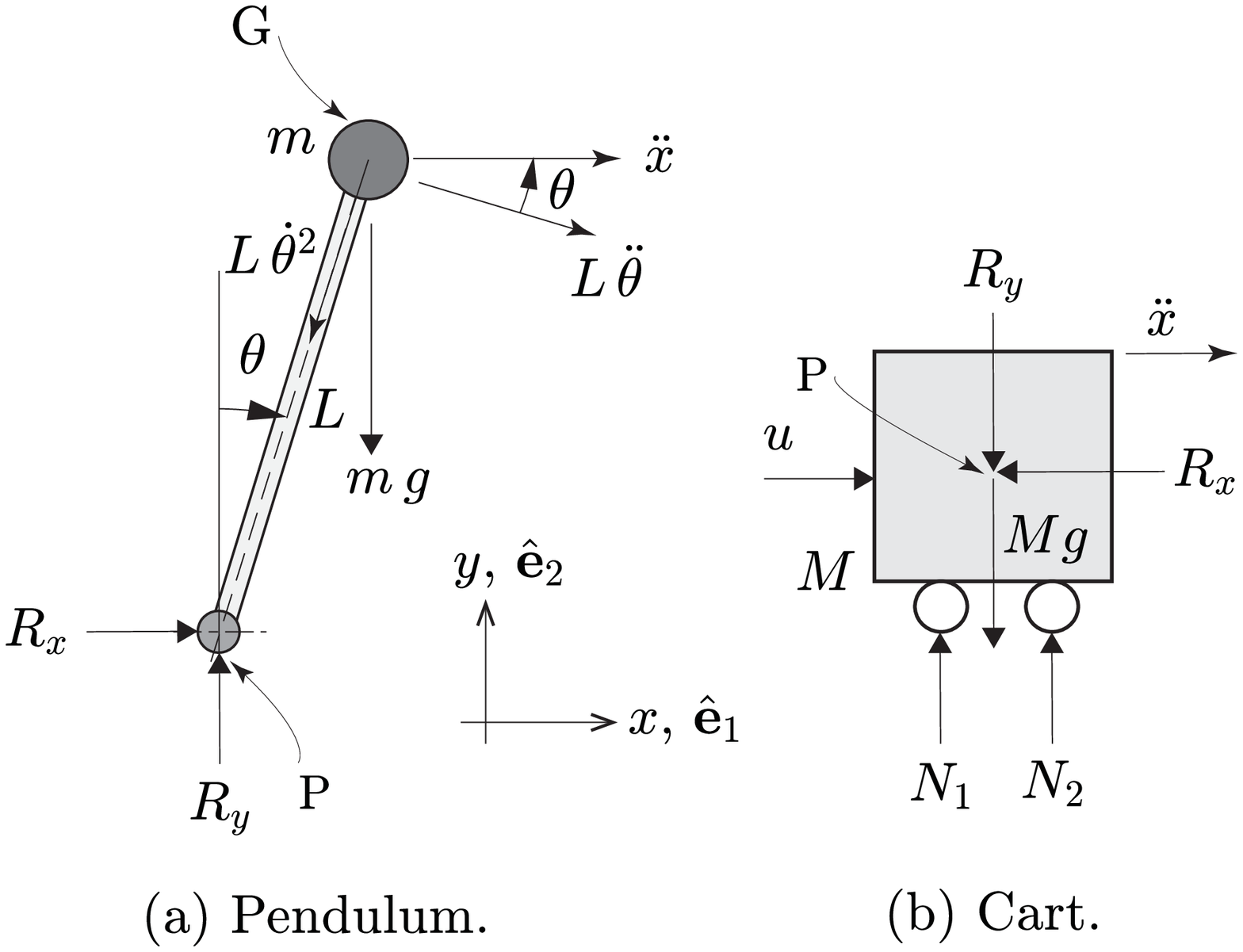}
	\caption{Free body diagrams for the pendulum and the cart.}
	\label{FBD}
\end{figure}

We draw free-body diagrams for both cart and pendulum (Fig.\ \ref{FBD}). The pendulum's pivot point P (see Fig.\ \ref{FBD}(a)) experiences a reaction force from the cart. This force has components \(R_x\)  and \(R_y\) along unit vectors \(\hat{\bf e}_1\) and \(\hat{\bf e}_2\) respectively. The cart experiences equal and opposite reactions. There are two ground forces on the cart wheels, named \(N_1\) and \(N_2\) (see Fig.\ \ref{FBD}(b)). Friction has not been included. The weight of the pendulum and cart are \(mg\) and \(Mg\) respectively. A horizontal force \(u\) acts on the cart.

Moving on to the kinematics, P has an acceleration \(\ddot x \,\hat{\bf e}_1\). The acceleration of G is
\begin{align}\label{acc_G}
	\boldsymbol{a}_G&=\boldsymbol{a}_P+\boldsymbol{a}_{G/P}\nonumber\\
	&=\ddot x\,\hat{\bf e}_1+\left(L\,\ddot \theta \,\hat{\bf e}_\theta-L\,{\dot \theta}^2\,\hat{\bf e}_n\right),
\end{align}
where \(\hat{\bf e}_\theta\) and \(\hat{\bf e}_n\) are unit vectors shown in Fig.\ \ref{FBD}(a).

For the pendulum, linear momentum balance gives
\begin{equation}\label{Rx_val}
R_x = m\left(\boldsymbol{a}_G\cdot\hat{\bf e}_1\right)= m\left( \ddot x  + L\,\ddot \theta \cos(\theta)-  L \,{\dot \theta}^2\sin(\theta)\right), 
\end{equation}
and for the cart, it gives
\begin{equation}\label{lmb_cart}
-R_x+ u= M\ddot{x}.
\end{equation}
Substituting \(R_x\) from Eq.\ (\ref{Rx_val}) in Eq.\ (\ref{lmb_cart})
\begin{equation}\label{lmb1}
M\ddot{x}+m\left(\ddot x+ L\,\ddot \theta\cos(\theta)\right)-m\,L\,{\dot \theta}^2\sin(\theta)=u.
\end{equation}
Now, for the pendulum, the moment about spatial point P (coincides with the pivot instantaneously) is
\begin{equation}\label{amb_eq}
\boldsymbol{\tau}_{\rm P}={\bf I}_{\rm G}\cdot\boldsymbol{\alpha}+\boldsymbol{r}_{\rm G/P}\times m\,\boldsymbol{a}_{\rm G},
\end{equation}
where \({\bf I}_{\rm G}\) is zero and \(\boldsymbol{r}_{\rm G/P}\) is the position vector of from P to G. This yields
\begin{equation}
-m\,g\,L\sin(\theta)=-m\,L\left(\ddot x\cos(\theta)+L\ddot \theta\right).	
\end{equation}
or
\begin{equation}\label{lmb2}
\ddot x\cos(\theta)+L\ddot \theta= g\sin(\theta).
\end{equation}
Linearizing Eqs.\ (\ref{lmb1}) and (\ref{lmb2}), for small \(\theta\) and \(\dot\theta\), we obtain
\begin{subequations}
\begin{equation}\label{lmb_x}
M\ddot{x}+m\left(\ddot x+ L\ddot \theta\right)=u,
\end{equation}
\begin{equation}\label{lmb_theta}
\ddot x+L\ddot \theta= g\theta.
\end{equation}
\end{subequations}

\section{Methodology used for obtaining \(C\) and \(P\)}
\label{method}
For both compensators \(C\) and \(P\), we choose \(n^{\rm th}\) order polynomials with unknown coefficients for both numerator and denominator, where \(n\) is a positive integer to be chosen by trial and error. The compensators' transfer functions are taken as
\[
C=\frac{a_0+a_1\,s+\dots a_n\,s^n}{1+a_{n+1}s+\dots+a_{2n}s^n},\, P=\frac{b_0+b_1\,s+\dots+b_n\, s^n}{1+b_{n+1}\,s+\dots+b_{2n}\,s^n},
\]
where \(a_k, b_k\) constitute \(4n+2\) unknown coefficients. 

Now we construct an objective function \(F\) as follows.
\begin{enumerate}
	\item[(i)] \(F\) takes \(4n+2\) numbers as a vector input \(\boldsymbol{q}\) and first forms the candidate \(C\) and \(P\).

	\item[(ii)] From the numerators and the denominators of \(C\) and \(P\), it calculates \(H\) (Eq.\ (\ref{CLTF})).

	\item[(iii)] It calculates the poles of \(C\), \(P\) and \(H\).

	\item[(iv)] From the poles of \(C\) and \(P\), the right-most real part is saved as \(p_1\).

	\item[(v)] From the poles \(z_{H_i}\) of \(H\), the right-most real part is saved as \(p_2\).

	\item[(vi)] A preliminary function value \(f_0\) is defined as
	\[f_0=\left\{\begin{array}{ll}
		p_2+6\,p_1 & \mbox{if $p_1\geq 0,$}\\
		p_2 & \mbox{otherwise},
	\end{array} \right.\]
   where the ``6'' is a penalty parameter, found to be big enough by trial and error (unnecessarily large penalty parameters are best avoided).

	\item[(vii)] For better behavior, the actual objective function used was
	\[
	F=f_0+\varepsilon_1\,\lvert\lvert\boldsymbol{q}\rvert\rvert+\varepsilon_2\,\max\{\lvert z_{H_i}\rvert\},\quad i=1,2,\dots,2\,n+4,
	\]
     where \(\varepsilon_1=10^{-5}\mbox{ and }\varepsilon_2=10^{-4}.\)
 
\end{enumerate}

If we can find a \(\boldsymbol{q}\) such that
\[ F(\boldsymbol{q})<0,\]
then our goal is accomplished.

We can now use any optimization techniques we like. We used a simple in-house genetic algorithm. The code is available on request.

\section{Controllability and observability}\label{app_modern_control}
So far, we have studied the system from the viewpoint of classical control theory. In the modern control approach, the system state consists of \(x,\,\theta,\,\dot{x}\) and \(\dot{\theta}\) given as a column matrix
\begin{equation}
	\boldsymbol{x}=\begin{Bmatrix}
		x\\\theta\\\dot x\\ \dot\theta
	\end{Bmatrix}.
\end{equation}
Writing Eqs.\ (\ref{eqm1}) and (\ref{eqm2}) in state space form, we obtain
\begin{equation}\label{state_space}
	\dot{\boldsymbol{x}}={\bf A}\,\boldsymbol{x}+\boldsymbol{B}\,u,
\end{equation}
where, for \(M=0.3\),
\begin{equation}
	{\bf A}=\left[ \begin {array}{cccc} 0&0&1&0\\ \noalign{\medskip}0&0&0&1
	\\ \noalign{\medskip}0&-{\frac{10}{3}}&0&0\\ \noalign{\medskip}0&{
		\frac{13}{3}}&0&0\end {array} \right]
	\quad
	\mbox{and}\quad
	\boldsymbol{B}= \begin {Bmatrix} 0\\ \noalign{\medskip}0\\ \noalign{\medskip}
	{\frac{10}{3}}\\ \noalign{\medskip}-{\frac{10}{3}}\end {Bmatrix}
	.
\end{equation}
In this problem, only the measurement of the cart displacement is available. So, the measured quantity
\begin{equation}
	y=x={\bf C}\,\boldsymbol{x},\quad\mbox{where }{\bf C}=[1\,\,\,\,0\,\,\,\,0\,\,\,\,0].
\end{equation}
Taking Laplace transforms of both sides of Eq.\ (\ref{state_space}), we obtain for zero initial conditions
\begin{equation}
	\boldsymbol{X}(s)=\left(s\,{\bf I}-{\bf A}\right)^{-1}\boldsymbol{B}\,U(s),  \quad \mbox{where }\boldsymbol{X}(s)=\mathcal{L}[\boldsymbol{x}(t)].
\end{equation}
Using the symbolic algebra package Maple, we have verified that
\begin{equation}
	{\bf C}\left(s\,{\bf I}-{\bf A}\right)^{-1}\boldsymbol{B}=G(s)=\frac{s^2-1}{s^2\,\left(0.3\,s^2-1.3\right)}.
\end{equation}
The  controllability matrix \cite{ogata2010modern} is
\begin{align}
	{\bf P}_{\rm C}&=\left[{\bf A}^3\boldsymbol{B}\,\lvert\,{\bf A}^2 \boldsymbol{B}\,\lvert\,{\bf A}\boldsymbol{B}\,\lvert\, \boldsymbol{B}\right]\nonumber\\
	&=
	\left[ \begin {array}{cccc} {\frac{100}{9}}&0&{\frac{10}{3}}&0
	\\ \noalign{\medskip}-{\frac{130}{9}}&0&-{\frac{10}{3}}&0
	\\ \noalign{\medskip}0&{\frac{100}{9}}&0&{\frac{10}{3}}
	\\ \noalign{\medskip}0&-{\frac{130}{9}}&0&-{\frac{10}{3}}\end {array}
	\right] ,
\end{align}  
which has full rank.
The observability matrix \cite{ogata2010modern}
\begin{equation}
	{\bf P}_{\rm O}=\begin{bmatrix}
		{\bf C}{\bf A}^3\\{\bf C}{\bf A}^2\\{\bf C}{\bf A}\\{\bf C}
	\end{bmatrix}=
	\left[ \begin {array}{cccc} 0&0&0&-{\frac{10}{3}}
	\\ \noalign{\medskip}0&-{\frac{10}{3}}&0&0\\ \noalign{\medskip}0&0&1&0
	\\ \noalign{\medskip}1&0&0&0\end {array} \right] 
\end{equation}
also has full rank. The system is both controllable and observable.
A controller can be designed by constructing a state estimator
and then using full state feedback. Let us consider the following system
\begin{subequations}
	\begin{equation}\label{est1}
		\dot{\boldsymbol{x}}={\bf A}\,\boldsymbol{x}-{\boldsymbol{B}}\,{\bf K}\,\tilde{\boldsymbol{x}}+{\boldsymbol{B}}\,u
	\end{equation}
	   \begin{equation}\label{est2}
	   	\dot{\tilde{\boldsymbol{x}}}={\bf A}\,\tilde{\boldsymbol{x}}-{\boldsymbol{B}}\,{\bf K}\,\tilde{\boldsymbol{x}}+{\boldsymbol{G}}\,{\bf C}\,(\boldsymbol{x}-\tilde{\boldsymbol{x}})+\boldsymbol{B}\, u
	   \end{equation}
\end{subequations} 
where \(\tilde{\boldsymbol{x}}\) is the estimated state and the gain matrices \({\bf K}\) and \({\boldsymbol{G}}\) are found by placing the system poles (arbitrarily) at
\begin{equation}\label{syspoles}
	-1\pm {\rm i}\mbox{ and }-2\pm {\rm i},
\end{equation}
and the estimator poles (also arbitrarily) at
\begin{equation}\label{estpoles}
	-1,-2,\,\mbox{ and }-3\pm {\rm i}
\end{equation}
on the complex plane. These numbers are chosen for demonstration only.
	
Combining Eqs.\ (\ref{est1}) and (\ref{est2}), we obtain
\begin{equation}\label{est_final}
	\dot{\tilde{\boldsymbol{Z}}}=\tilde{\bf A}\,\tilde{\boldsymbol{Z}}+\tilde{\boldsymbol{B}}\,u,
\end{equation}
where, 
\begin{equation}
    \tilde{\boldsymbol{Z}}=\begin{Bmatrix}
    	\boldsymbol{x}\\ \tilde{\boldsymbol{x}}
    \end{Bmatrix},
    \,
    \tilde{\bf A}=\left[\begin{array}{cc}
    	{\bf A}&-{\boldsymbol{B}}\,{\bf K}\\
    	{\boldsymbol{G}}\,{\bf C}&{\bf A}-{\boldsymbol{B}}\,{\bf K}-{\boldsymbol{G}}\,{\bf C}
    \end{array}\right],\,\mbox{ and }\,
    \tilde{\boldsymbol{B}}=\begin{Bmatrix}
    	{\boldsymbol{B}}\\{\boldsymbol{B}}
    \end{Bmatrix}.
\end{equation}
The output 
\begin{equation}
    y={\bf C}\,\boldsymbol{x}=\tilde{\bf C}\,\tilde{\boldsymbol{Z}},\quad\mbox{where  }\tilde{\bf C}=[{\bf C},0,0,0,0].
\end{equation}
To interpret these result in light of the main paper, we can now think of an implied feedback controller, with closed loop transfer function
\begin{equation}
	Q(s)=\tilde{\bf C}\left(s{\bf I}-\tilde{\bf A}\right)^{-1}\tilde{\boldsymbol{B}}. 
\end{equation}
Using Maple, we obtain
\begin{equation}
	  Q(s)={\frac {10\,{s}^{2}-10}{3\,{s}^{4}+18\,{s}^{3}+45\,{s}^{2}+54\,s+30}}.
\end{equation}
We observe that \(Q\) and \(G\) share the same zeros, and the poles of \(Q\) are the same as the system poles chosen for placement (Eq.\ (\ref{syspoles})).
\begin{figure}[h!]
  \centering
  \includegraphics[scale=0.6]{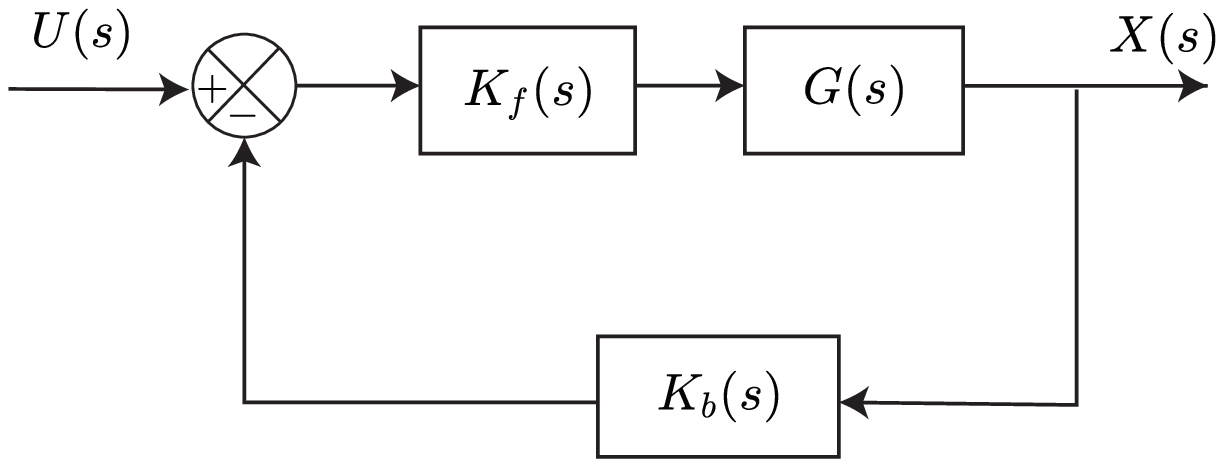}
  \caption{An equivalent single loop feedback control system.}
  \label{equiv_sys}
\end{figure}
We may think of a feedback control system equivalent to the implied control system as shown in Fig.\ \ref{equiv_sys}, where plant \(G\) is assigned compensators \(K_b\) and \(K_f\) in forward and feedback loops respectively.
Hence
\begin{equation}
    Q=\frac{K_f\,G}{1+K_b\,K_f\, G}.
\end{equation}
From algebraic manipulations, we obtain;
\begin{equation}
   K_f=\frac{Q}{G}\,\frac{1}{1-K_b \,Q} \quad\mbox{and}\quad K_b=\frac{1}{Q}-\frac{1}{K_f\,G}.
\end{equation}
Clearly, there are infinitely many solutions for \(K_f\) and \(K_b\). We examine two limiting cases for better understanding. 
\begin{enumerate}
	
	\item [(i)] \emph{The system shown in Fig.\ \ref{equiv_sys} has a compensator only in the feedback loop}, i.e., \(K_f=1\). In this case,
	\begin{equation}\label{Kb_exp}
		K_b=\frac{1}{Q}-\frac{1}{G}=\frac {9\,{s}^{3}+29\,{s}^{2}+27\,s+15}{5\,{s}^{2}-5},
	\end{equation}
	which is unacceptable (both improper and unstable).
	
	\item[(ii)] \emph{The system shown in Fig.\ \ref{equiv_sys} has a compensator only in the forward loop}, i.e., \(K_b=1\). Now we have
	\begin{align}\label{Kf_exp}
		K_f&=\frac{Q}{G}\,\frac{1}{1-Q}=\frac{d_G}{d_Q-n_Q}\nonumber \\
		&={\frac {{s}^{2} \left( 3\,{s}^{2}-13 \right) }{3\,{s}^{4}+18\,{s}^{3}+
				35\,{s}^{2}+54\,s+40}},
	\end{align}
	where \(d_Q\) and \(n_Q\) are the denominator and numerator of \(Q\) respectively. 
	
	The compensator \(K_f\) is stable, but relies on pole zero cancellation which is not allowed in classical control. A commonly stated reason for not allowing pole zero cancellation is that the slightest inaccuracy in the controller will destroy the cancellation and instability will reappear. Youla et al. \cite{youla1974single} also point out that exact pole zero cancellation may represent nonobservable modes which remain unstable. In any case, we cannot accept this \(K_f\). 
	
\end{enumerate}

We already know that the controller obtained in this appendix cannot be realized (Youla et al. \cite{youla1974single}) with stable and proper compensators in the classical single loop configuration. Equations (\ref{Kb_exp}) and (\ref{Kf_exp}) merely provide two examples of the difficulties encountered if such an attempt is made.

\section{Robustness and fragility}\label{robustness}

Having found a stable closed loop transfer function \(H\) as explained in appendix \ref{method}, we can check its sensitivity to small changes in plant and compensator parameters.

\begin{figure}[h!]
\centering
\includegraphics[width=\linewidth]{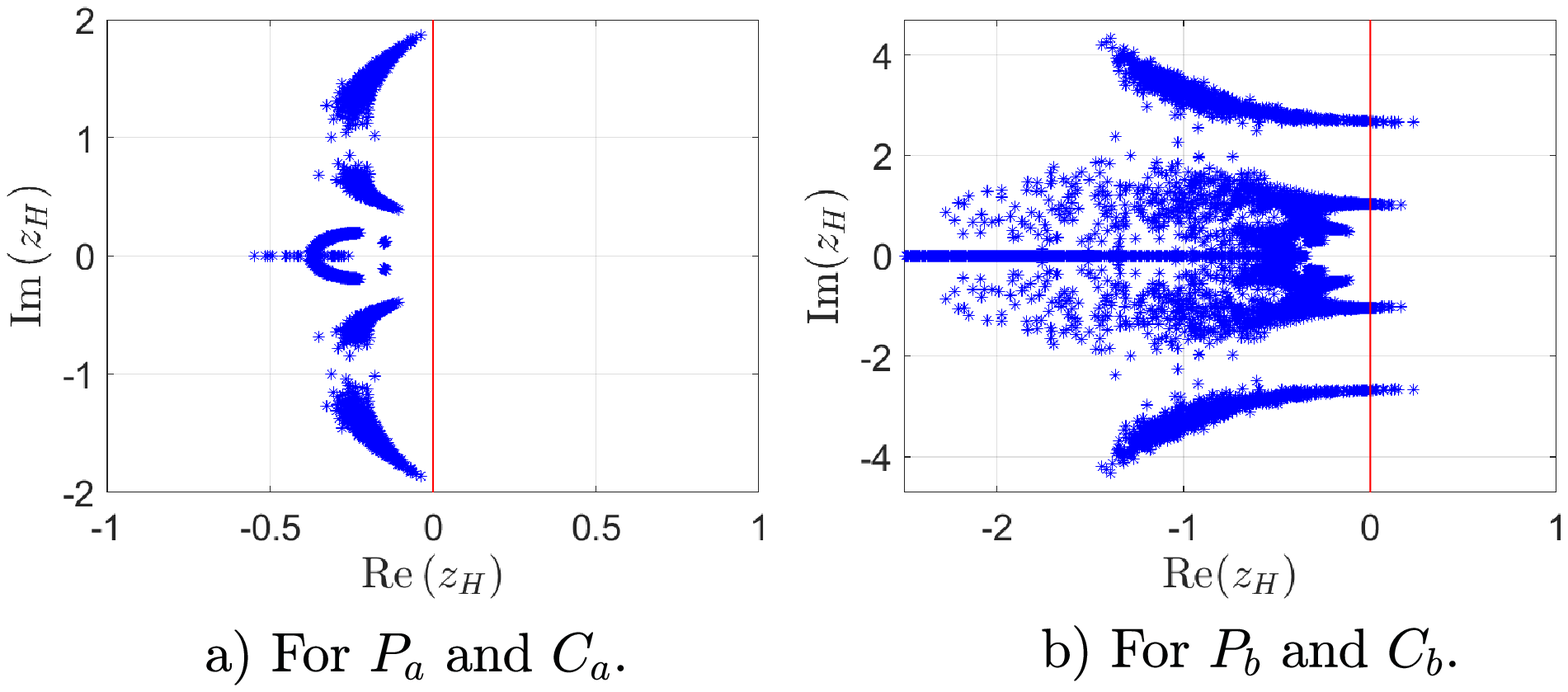}
\caption{Robustness under small changes in plant parameters (1000 random perturbations).}
\label{robustness_fig}
\end{figure}

Robustness, for a control system, is its ability to retain stability under small changes in the plant parameters. Here, the plant parameters depend on the system parameters: \(L,\,m,\,g\) and \(M\). Of them, the first three were eliminated from the governing equations by introducing nondimensional displacement, time and mass \(\tilde{x},\,\tilde{t}\) and \(\tilde{m}\) respectively where
\begin{equation}\label{nondim_eq}
\tilde{x}=\frac{x}{L},\quad\tilde{t}=t\sqrt{\frac{g}{L}},\quad\tilde{m}=\frac{M}{m},
\end{equation}
 which is analogous to setting the values of \(m,\,L\) and \(g\) to unity and treating \(M\) as the only free parameter. So far we have considered \(M=0.3\). To investigate the effect of small changes in parameter values on the system behavior, we rewrite the plant transfer function as
\begin{equation}
G=\frac{A_0\, s^2-A_1}{s^2\,\left(0.3\,A_2\,s^2-1.3\,A_3\right)},
\end{equation}
where the parameters \(A_0,\,A_1,\,A_2\) and \(A_3\) are notionally equal to unity along with \(M=0.3\). For a large number of random calculations (1000 times), we perturb the $A$'s by normally distributed \emph{iid} random variables \(r_i,\,i=0,1,2,3\), with zero mean and standard deviation 0.02 (99.7\% of the points are within $\pm$ 6\%) in the following way
\[A_i\mapsto A_i\,(1+r_i),\quad i=0,1,2,3,.\]
We then plot the poles (\(z_H\)) of the respective closed loop transfer functions (CLTF). Results are shown in Fig.\ \ref{robustness_fig}.

For compensators \(C_a\) and \(P_a\), the entire cloud (Fig.\ \ref{robustness_fig}(a)) of poles remains in the left half plane. For compensators \(C_b\) and \(P_b\), a significant part of the cloud (Fig.\ \ref{robustness_fig}(b)) remains in the left half plane. In 44 out of 1000 cases, the CLTF has poles in the right half plane. Thus, the compensators are fairly robust; and \(C_a\) and \(P_a\) are more robust than \(C_b\) and \(P_b\).

\begin{figure}[h!]
	\centering
	\includegraphics[width=\linewidth]{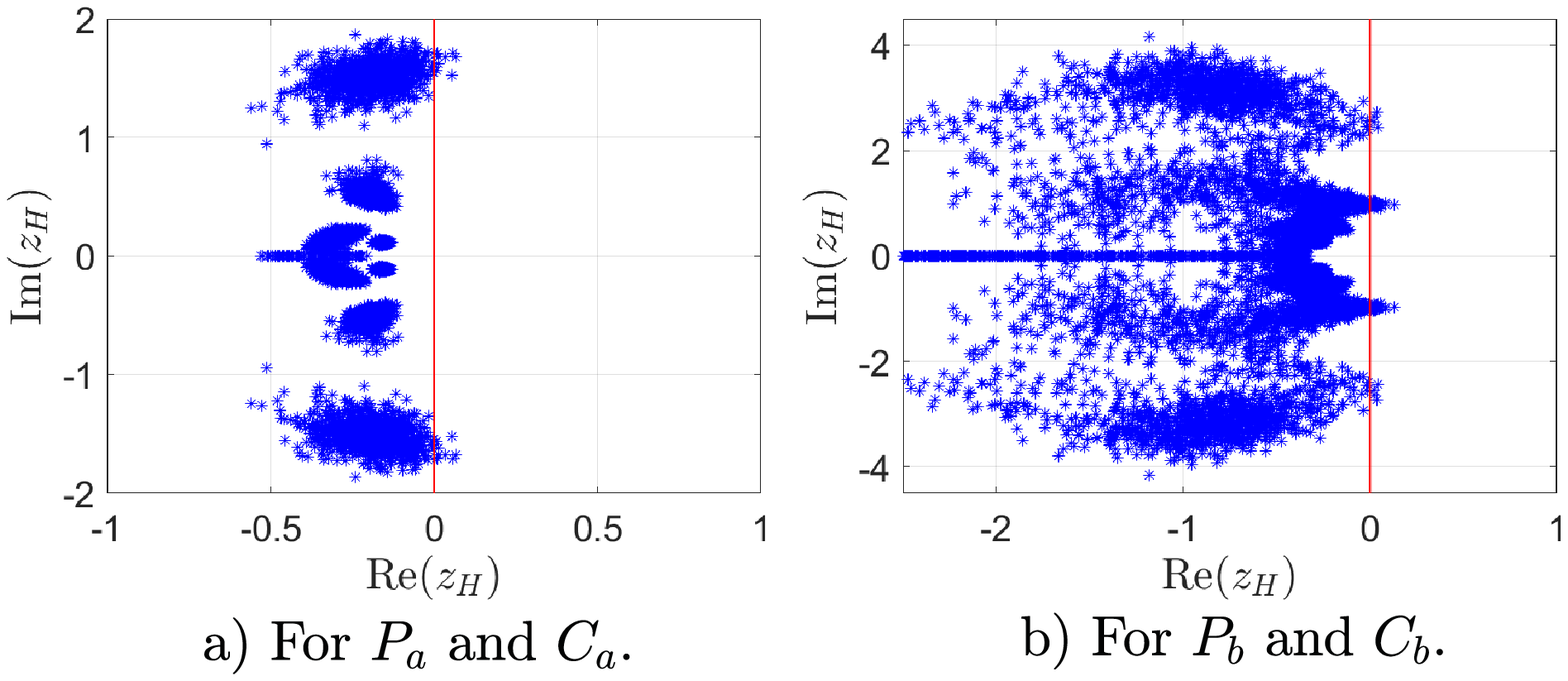}
	\caption{Performance of the system under small changes in compensator parameters (1000 random perturbations).}
	\label{fragility_fig}
\end{figure}

Some robust control systems perform poorly under small perturbations in the compensator parameters. This is called the {\em fragility} \cite{keel1997robust} of the system. To check fragility, we perturb the compensator parameters, again 1000 times, by normally distributed \emph{iid} random variables \(s_i,\,i=0,1,\dots,4\,n+1\). Here, \(n\) is degree of the polynomials in the numerator and denominator of the compensators. The random variables \(s_i\) have zero mean and standard deviation 0.02 (99.7\% of them are within $\pm$ 6\%). We perturb the \(a\)'s and \(b\)'s as follows:
\[a_i\mapsto a_i\,(1+s_i),\quad b_i\mapsto b_i\,(1+s_{2\,n+1+i}),\quad i=0,1,\dots,2\,n.\]
We then calculate the poles (\(z_H\)) of the respective closed loop transfer functions. The results are shown in Fig.\ \ref{fragility_fig}.
For the compensators \(C_a\) and \(P_a\), a large portion of the cloud of poles again remains in the left half plane. In 9 out of 1000 cases, the CLTF has poles in the right half plane. For the compensators \(C_b\) and \(P_b\), in 32 out of 1000 cases, the CLTF has poles in the right half plane.

We conclude with the following observation. 
Implementability, albeit implicitly discussed, has motivated this entire paper. Finding stable compensators (which we have now shown are fairly robust and not fragile) indicates that the compensators are implementable.

\section{Response to noise}\label{app_noise}
In section \ref{effect_of_noise}, we examined the system's sensitivity to six noise inputs \(e_i(t),\,i=1,2,\dots,6\) by using Bode plots.
To demonstrate the effect of noise on the time response of the system, we use the following input
\begin{equation}
e_i(t)=\sum_{k=1}^{N}c_k\sin\left(\omega_k\, t\right),\quad i=1,2,\dots,6,
\end{equation}
\begin{figure}[h!]
	\centering
	\includegraphics[scale=0.5]{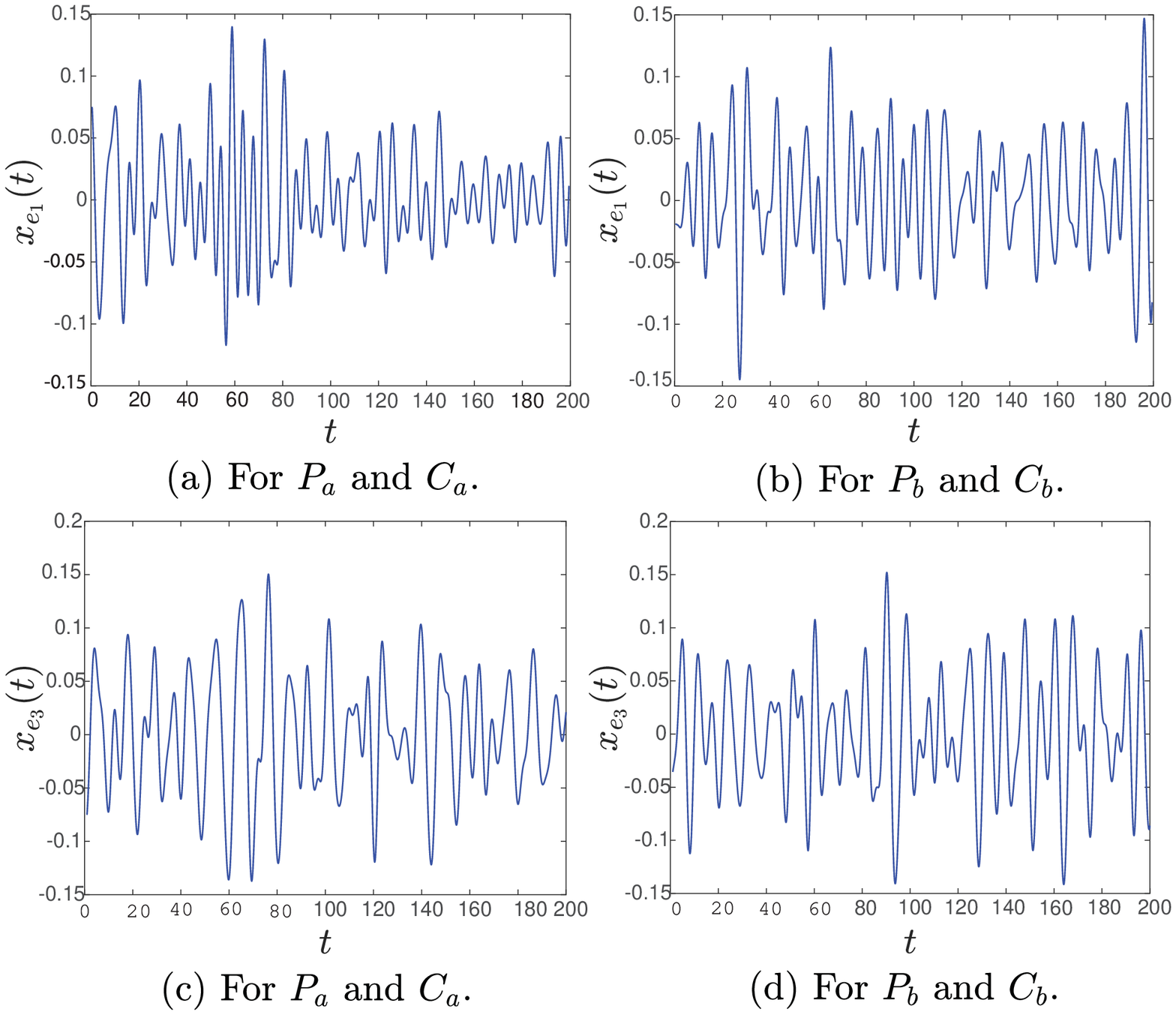}
	\caption{Time responses to noise inputs \(e_1(t)\) and \(e_3(t)\).}
	\label{noise_fig}
\end{figure}
where the \(\omega\)'s are randomly chosen numbers uniformly distributed in the interval \([0.5,1.5]\). The amplitudes \(c_k,k=1,\dots,N\) are random numbers where the norm of the vector \(\boldsymbol{c}=[c_1,\,c_2,\dots,c_N]^\top\) is set to 0.01. For calculations, we have used \(N=4000\). The response, with phase randomized, is taken as
\begin{multline}
x_{e_i}(t)=\sum_{k=1}^{N}c_k \lvert H_{e_i}\!\left({\rm i}\,\omega_k\,\right)\rvert\,\sin\left(\omega_k\, t+\arg\left(H_{e_i}({\rm i}\,\omega_k)\,\right)+\phi_k\right),\\ i=1,2,\dots,6,
\end{multline}
where \({\rm i}=\sqrt{-1}\), and the \(\phi_k\) are random numbers uniformly distributed in the interval \([0,2\,\pi]\).

In Fig.\ \ref{noise_fig}, the amplification factor is consistent with the Bode plots of section \ref{effect_of_noise}.

\small{}

\end{document}